\documentclass[12pt]{iopart}

\usepackage{iopams}
\usepackage{bbm}
\usepackage{graphicx}

\newcommand{\Kast}{K^\ast}
\newcommand{\arccosh}{\mbox{arccosh}}
\newcommand{\dd}{\mathrm{d}}
\newcommand{\fsolv}{f_\mathrm{solv}}
\newcommand{\Tast}{T^\ast}

\newcommand{\TwM}{T_{\mathrm{w},M}}
\newcommand{\TcM}{T_{\mathrm{c},M}}
\newcommand{\Tw}{T_\mathrm{w}}
\newcommand{\Tc}{T_\mathrm{c}}
\newcommand{\sign}{\mbox{sign}}
\newcommand{\TwMG}{T_{\mathrm{w},M}^\gamma}
\newcommand{\TcMG}{T_{\mathrm{c},M}^\gamma}
\newcommand{\kb}{k_\mathrm{B}}

\begin{document}
\title{Properties of the solvation force of a~two-dimensional Ising
  strip in scaling regimes}
\author{Piotr Nowakowski and Marek Napi\'orkowski}
\address{Institute of Theoretical Physics, University of Warsaw, Ho\.za 69, 00-681 Warszawa, Poland}
\ead{pionow@fuw.edu.pl}
\begin{abstract} 
We consider $d=2$ Ising strip with surface fields acting on boundary spins. Using the properties of the transfer matrix spectrum we identify two pseudotransition temperatures and show that they satisfy similar scaling relations as expected for real transition temperatures in strips with $d>2$. The solvation force between the boundaries of the strip is analysed as a function of temperature, surface fields and the width of the strip. For large widths the solvation force can be described by scaling functions in three different regimes: in the vicinity of the critical wetting temperature of 2D semi-infinite system, in the vicinity of the bulk critical temperature, and in the regime of weak surface fields where the critical wetting  temperature tends towards the bulk critical temperature. The properties of the relevant scaling functions are discussed.
\end{abstract}

\section{Introduction}

Fluctuating condensed-matter systems enclosed by walls are characterised by the appearance of solvation force acting between the walls. This force originates from the fluctuations of the confined system. The properties of solvation forces have been the subject of increasing interest during the last years \cite{T1,D2,T3,T4,danchev,T5,D6,kardar,T7,T8,D9,D10,D11,T12,D13,D14,T15,NN}. Both the  shapes and possible chemical inhomogeneity of the confining walls influence the form of solvation forces \cite{T4,T5,D13,D14,T15} which additionally depend on the thermodynamic state of the system and on the interaction between the system and the walls. In particular, if the system is chosen to be at its bulk critical point the solvation forces become long ranged and show universality \cite{krech} while in the vicinity of criticality scaling behaviour is observed \cite{pe1,barber}.  

In this article we analyse the solvation forces in two-dimensional Ising strips. The spins are confined by two parallel, planar and chemically homogeneous walls separated by distance $M$. Each wall interacts with the system by surface fields acting on the boundary spins. Our goal is the exact determination of the properties of the solvation forces as functions of temperature, surface fields and the width of the strip $M$. To investigate these properties we a use method based on exact diagonalization of the transfer matrix which is followed by numerical determination of appropriate eigenvalues.

The paper is organised as follows. In \sref{sec1} we define the 2D Ising strip, recall its  properties in the bulk limit as well as the properties of the semi-infinite system related to critical wetting. In \sref{sec2} we define pseudotransition temperatures and check that they 
display the scaling properties expected for higher dimensional systems. \Sref{sec3} is devoted to our main goal, i.e. analysis of the properties of the solvation force acting between the system boundaries. We first recall the definition of solvation force and adapt it to our model. We study this force numerically to establish several of its properties as functions of temperature, surface fields and the distance between the walls. For large width of the strip we explain these properties by introducing scaling functions in three different scaling regimes: around the wetting temperature, around the bulk critical temperature, and in regime in which both of the above temperatures are close to each other.


\section{Ising strip}\label{sec1}
\subsection{The model}

We consider an Ising model on a two-dimensional square lattice with $N$ columns and $M$ rows, and impose periodic boundary conditions in the horizontal direction. In this way the Ising strip of width $M$ is obtained. We assume that surface fields $h_1$ and $h_2$ act on the spins located at the bottom and the top row, respectively; these fields can be considered as model short range interactions between the system and the surrounding walls. The Hamiltonian of the system has the form
\begin{equation}
\fl{\cal H}\left(\left\{ s_{n,m}\right\}\right) =-J \sum_{n=1}^N \sum_{m=1}^{M-1} (s_{n,m}s_{n+1,m}+s_{n,m}s_{n,m+1})-\sum_{n=1}^N (h_1 s_{n,1} + h_2 s_{n,M}),
\end{equation}
where $s_{n,m}=\pm 1$ denotes the spin located in the $n$-th column and $m$-th row, and $s_{N+1,m}=s_{1,m}$. The coupling constant $J$ is positive (ferromagnetic case) and we assume no bulk field $h$ acting on the system.  

In this paper, we concentrate on two special choices of surface fields corresponding to the so-called symmetric and antisymmetric case: in the symmetric case (denoted by the superscript S) one has $h_1=h_2$ while in the antisymmetric case (AS) $h_1=-h_2$. Later on, we will also use superscript O to denote the limiting case $h_1=h_2=0$ which is referred to as the free case.

\subsection{The free energy of the strip}

To calculate the free energy of our system we use the method based on
exact diagonalization of the transfer matrix. In this method the
$2^{M+2}\times 2^{M+2}$ transfer matrix is represented by
$\left(2M+4\right)\times\left(2M+4\right)$ orthogonal matrix
$R$. Eigenvalues of the transfer matrix are calculated from
eigenvalues of $R$ which can be found by solving recurrence equations for eigenvectors of $R$ \cite{kaufman, abraham, maciolek}. Here we only recall the final formulae for the free energy per column (here and in the following formulae we do not explicitly write the dependence of the free energy  and other quantities on the coupling constant $J$)
\begin{eqnarray}
\fl\bar{f}^\mathrm{S}\left(T,h_1,M\right) &=&-\kb T\left[ \frac{1}{2}\left( \gamma_1+\gamma_2+\gamma_3+\ldots+\gamma_{M+1}\right)+\frac{M}{2}\ln\left(2
    \sinh 2 K\right)\right],\label{free:e1}\\
\fl\bar{f}^\mathrm{AS}\left(T,h_1,M\right)&=&-\kb T\left[ \frac{1}{2}\left( -\gamma_1+\gamma_2+\gamma_3+\ldots
    +\gamma_{M+1}\right)+\frac{M}{2} \ln\left( 2\sinh 2K \right)\right],\label{free:e2}
\end{eqnarray}
where $K=J/\kb T$ and $\kb$ is the Boltzmann constant.
The coefficients $\gamma_1<\gamma_2<\ldots<\gamma_{M+1}$ are positive functions of parameters $T, h_1$ and $M$ defined by relation
\begin{equation}\label{omega2gamma}
\cosh \gamma_k=\cosh \left(2K-2\Kast\right)+1-\cos\omega_k,
\end{equation}
where parameter $\Kast$ is obtained from $\sinh 2K \sinh 2\Kast=1$. The functions $\omega_k$ are solutions of the equations
\begin{equation}\label{eq:omega}
\left(M+1\right)\omega_k-\delta^\prime\left(\omega_k,T\right)-\phi\left(\omega_k,T,h_1\right)=\left(k-l\right) \pi,\qquad 0<\omega_k<\pi,
\end{equation} 
and $k=1,2,\ldots,M+1$. The function $l\left(T,h_1\right)$ is defined as
\begin{equation}\label{l}
l\left(T,h_1\right)=\left\{\begin{array}{rl} 2 & \mbox{for } T<\Tw, \cr 0 & \mbox{for } \Tw<T<\Tc, \cr  1 & \mbox{for } T>\Tc. \end{array}\right.
\end{equation}
The symbol $\Tc$ denotes the bulk critical temperature \cite{KW} 
\begin{equation}
K_\mathrm{c}=\frac{J}{\kb\Tc}=\frac{1}{2}\ln\left(1+\sqrt{2}\right),
\end{equation}
while $\Tw\left(h_1\right)$ denotes the temperature of the critical wetting transition taking place in the semi-infinite Ising model. It depends on the surface field $h_1$ and can be defined by 
equation \cite{ms}
\begin{equation} \label{Tw}
W\left(\Tw,h_1\right)=1,
\end{equation}
where
\begin{equation}\label{W}
W\left(T,h_1\right)=\left(\cosh 2\Kast+1\right)\left(\cosh 2K-\cosh 2 K_1\right), \qquad  
K_1=\frac{h_1}{\kb T}. 
\end{equation}
We observe that for certain ranges of temperatures \eref{eq:omega} may not have a solution for $k=1$ and $k=2$. In such cases $\omega_1$ and $\omega_2$ are imaginary and satisfy equations
\begin{equation}\label{eq:omega1}
\fl\omega_k=\rmi u_k,\qquad \rme^{-u_k M}=\alpha_k \exp\left\{\rmi\left[\phi\left(\rmi u_k,T\right)+\delta^\prime\left(\rmi u_k, T, h_1\right)\right]\right\}, \qquad k=1,2,
\end{equation}
with $\alpha_k=\pm 1$. The detailed rules for selecting the signs of $\alpha_1$ and $\alpha_2$ are presented in the next subsection. Functions $\phi$ and $\delta^\prime$ are calculated from the formulae
\begin{equation}\label{phi:delta}
\rme^{\rmi \phi\left(\omega, T,h_1\right)}=\rme^{\rmi \omega} \frac{W \rme^{\rmi \omega}-1}{\rme^{\rmi\omega}-W},\qquad
\rme^{2 \rmi \delta^\prime\left(\omega,T\right)}=\frac{\left(\rme^{\rmi\omega}-A\right)\left(\rme^{\rmi\omega}-B\right)}{\left( A\rme^{\rmi\omega}-1\right)\left(B\rme^{\rmi\omega}-1\right)},
\end{equation}
where $A\left(T\right)=\left(\tanh K\tanh \Kast\right)^{-1}$, $B\left(T\right)=\tanh K/\tanh \Kast$, and the function $W\left(T,h_1\right)$ is given in \eref{W}.
To determine the angles $\phi\left(\omega,T,h_1\right)$ and
$\delta^\prime\left(\omega,T\right)$ uniquely we pick the continuous branches of solutions for which 
\begin{equation}
\phi\left(0,T,h_1\right)=\pi, \qquad \delta^\prime\left(0,T\right)=-\pi.
\end{equation}
For $\omega=0$ and $T=\Tw$, the angle $\phi\left(\omega,T,h_1\right)$ is undefined ($W\left(\Tw,h_1\right)=1$) while at $T=\Tc$ the angle $\delta^\prime\left(\omega,T\right)$ is undefined ($K=\Kast$, so $B=1$). Although at these temperatures our formulae are useless one can use the continuity of the free energy and calculate it using a limiting procedure.

\subsection{The characteristic temperatures}

Because the lower critical dimension of the Ising model equals two
($d_l=2$), no true transition may occur in a two-dimensional Ising strip
with finite $M$. On the other hand, the infinite 2D Ising model
experiences the critical point behaviour at $T=\Tc$, while in the semi-infinite 2D Ising model with the surface field $h_1$, the critical wetting transition takes place at $T=\Tw$, $\Tw<\Tc$ \cite{pe1}. Below, we discuss the properties of \eref{eq:omega} and on this basis we define the characteristic temperatures $\TwMG$ and $\TcMG$.

First we consider $T<\Tw$ case, for which $l=2$, see \eref{l}. For small enough temperatures the left-hand side of \eref{eq:omega} is an increasing function of $\omega$, it equals 0 for $\omega=0$, and thus this equation does not have a solution for $k=1$ and $k=2$. The coefficients $\omega_1$ and $\omega_2$ are thus found from \eref{eq:omega1} with $\alpha_1=-1$ and $\alpha_2=1$. However, when $T$ is getting close to $\Tw$ the situation becomes different: the left-hand side of \eref{eq:omega} --- upon increasing $\omega$ --- first decreases, has a minimum and then increases. As a result \eref{eq:omega} has a solution for $k=2$. At the same time, to obtain coefficient $\omega_1$ equation \eref{eq:omega1} must be used with $\alpha_1=-1$. The $M$-dependent temperature, which separates the above two possibilities is denoted by $\TwMG$.

When $\Tw<T<\Tc$, one has $l=0$ and all coefficients $\omega_k$ are defined by \eref{eq:omega}.

For $T>\Tc$, $l=1$ and for temperatures well above $\Tc$ equation \eref{eq:omega} does not have solution for $k=1$; the coefficient $\omega_1$ can be calculated from \eref{eq:omega1} with $\alpha_1=+1$. When $T$ is close to $\Tc$ the left-hand side of \eref{eq:omega} is a non-monotonic function of $\omega$, and thus the solution exists for any $k$. The characteristic temperature separating these two cases is denoted by $\TcMG$.

\begin{figure}
\begin{center}
\begin{tabular}{ll}
(a) & (b) \\
\includegraphics[width=0.45\textwidth]{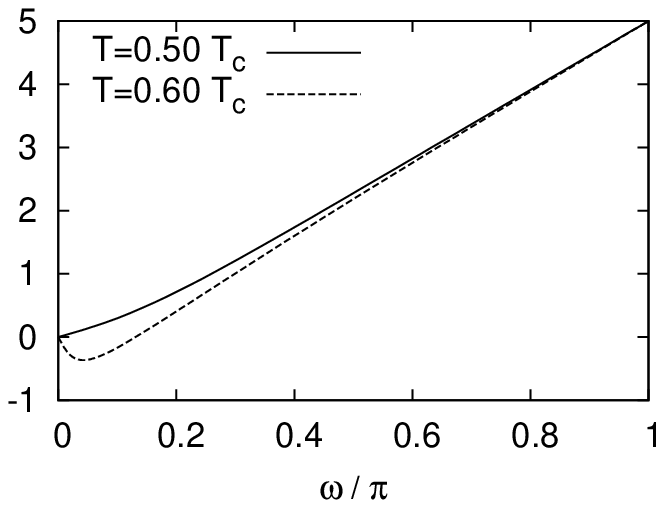} & 
\includegraphics[width=0.45\textwidth]{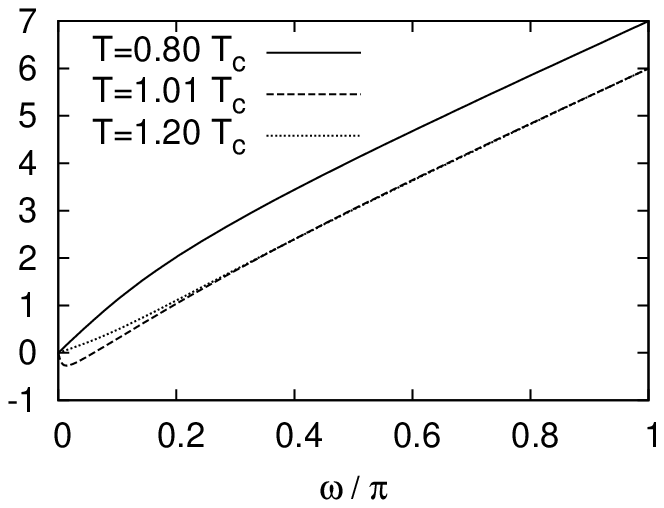} \\
\end{tabular}
\end{center}
\caption{\label{omega} Plots of the left-hand side of \eref{eq:omega} divided by $\pi$  for $h_1=0.8 J$ ($\Tw \approx 0.621\ \Tc$) and $M=5$. The solutions $\omega_k$ correspond to integer values of this function. For $T=0.5 \Tc<\TwMG$ there are $M-1=4$ solutions, for $\TwMG<T=0.6\Tc<\Tw$ there are $M=5$ solutions, for $\Tw<T=0.8\Tc<\Tc$ there are $M+1=6$ solutions, for $\Tc<T=1.01\Tc<\TcMG$ there are $M+1=6$ solutions, and for $\TcMG<T=1.2\Tc$ there are $M=5$ solutions. Total number of solutions is $M+1=6$; the missing solutions correspond to imaginary values of $\omega$ and are determined from \eref{eq:omega1}. } 
\end{figure}

Typical plots of the left-hand side of \eref{eq:omega} are shown in \fref{omega}.

\begin{figure}
\begin{center}
\includegraphics[width=0.45\textwidth]{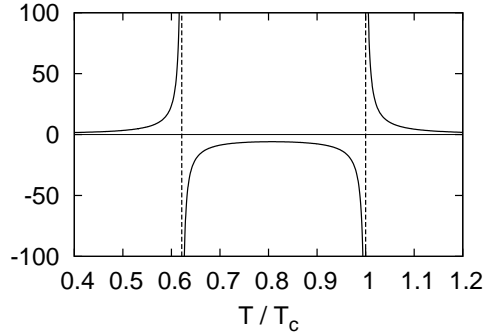}
\end{center}
\caption{\label{eq15}Plot of the left-hand side of \eref{TwM:TcM} for $h_1=0.8 J$ (for which $\Tw\approx 0.621\ \Tc$). At pseudotransition temperatures $\TwMG\left(h_1,M\right)$ and $\TcMG\left(h_1,M\right)$ this function is equal to $M+1$. To guide an eye the vertical broken lines corresponding to $T=\Tw$ and $T=\Tc$ are drawn.}
\end{figure}

To find the formulae for $M$-dependent temperatures $\TwMG$ and $\TcMG$ we use the fact that at these two temperatures \eref{eq:omega}  has double solution for $\omega=0$. In other words, the condition
\begin{equation}
\left. \frac{\partial}{\partial \omega}\right|_{\omega=0}\left[\left( M+1\right)\omega-\delta^\prime\left(\omega, T\right)-\phi\left(\omega,T,h_1\right)\right]=0
\end{equation}
must be satisfied, which leads to
\begin{equation}\label{TwM:TcM}
\frac{2W\left(T,h_1\right)}{W\left(T,h_1\right)-1}-\frac{\sinh 2K}{\sinh\left(2K-2\Kast\right)}=M+1,
\end{equation}
where $W\left(T,h_1\right)$ is defined in \eref{W}. To find solutions
of this equation it is useful to analyse its left-hand side as a
function of temperature: it equals to $1$ for $T=0$, is an
increasing function of temperature for $0<T<\Tw$, at $T=\Tw$ 
reaches infinity and has a pole ($W=1$ for $T=\Tw$). For $\Tw<T<\Tc$
the left-hand side of \eref{TwM:TcM} is negative and has another pole
for $T=\Tc$ ($K=\Kast$ at $T=\Tc$). For $T>\Tc$ it decreases from infinity at $T=\Tc$ to $0$ for $T\to\infty$. A typical plot of left-hand side of \eref{TwM:TcM} is shown in \fref{eq15}. \Eref{TwM:TcM} has two solutions for any positive $M$ -- the solution $\TwMG$ is always smaller that $\Tw$ and approaches the wetting temperature monotonically as $M\to\infty$, while the solution $\TcMG\left(h_1,M\right)$ is always larger than $\Tc$ and decreases monotonically to $\Tc$ as $M\to\infty$.

\section{Properties of pseudotransition temperatures}\label{sec2}

In an infinite strip of width $M$ and dimension $d$ larger than the
lower critical dimension $d_l$, $d>d_l=2$, true phase transitions
corresponding to the non-analyticity of free energy occur.

In a strip with symmetric surface fields (S) capillary condensation is
expected. For a vanishing bulk field the strip is filled with phase favoured by the walls for $T<\TcM$. The critical temperature $\TcM$ is shifted away from $\Tc$ \cite{EMT}. On the other hand, for antisymmetric surface fields  (AS) with no bulk field, a transition is observed at $\TwM$ that is shifted from $\Tw$. For $T<\TwM$ the interface separating two phases is located close to one of the walls while for $T>\TwM$ this interface is located in the middle of the system \cite{pe1}. Temperature $\TwM$ approaches $\Tw$ as $M\to\infty$. In AS case the second phase transition at $\TcM$ located close to $\Tc$, also occurs.

In a two-dimensional strip no phase transition may occur for finite $M$. However, for large strip widths we expect some thermodynamics functions to vary rapidly close to certain temperature values while remaining analytic. It is convenient to define these pseudotransition temperatures which can be then used to characterise the behaviour of our system. Since all functions are analytic, these temperatures cannot be defined uniquely. There are different criteria according to which the  pseudotransition temperature can be defined and thus there is no single $\TwM$ and $\TcM$. One possibility corresponds to $\TwM^\mathrm{c}$ and $\TcM^\mathrm{c}$ defined as the temperatures at which the specific heat attains its maximum values. Here we would like to show that the just defined temperatures $\TwMG$ and $\TcMG$ may be treated as such  pseudotransition temperatures.

\begin{figure}
\begin{center}
\includegraphics[width=0.45\textwidth]{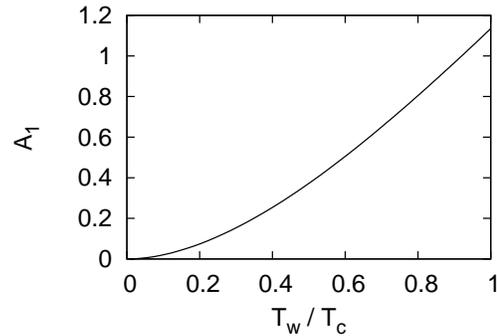}
\end{center}
\caption{\label{fig1}The plot of the amplitude $A_1$ (see \eref{amplitude:A1}) as a function of $\Tw$.}
\end{figure}

First we check how the difference $\Tw-\TwMG$ depends on the width of the strip $M$ for large $M$. This can be done on the basis of \eref{TwM:TcM}. Using the implicit function theorem one obtains 
\begin{equation}\label{amplitude:A1}
\fl \frac{\Tw-\TwMG}{\Tc}=\frac{A_1\left(h_1\right)}{M}+\Or\left(M^{-2}\right), \qquad A_1\left(h_1\right)=2\left(\Tc \left.\frac{\partial W}{\partial T}\right|_{T=\Tw} \right)^{-1}.
\end{equation}
\Fref{fig1} shows the plot of the amplitude $A_1\left(\Tw\right)$ after reparametrization from $h_1$ to $\Tw$ has been done according to \eref{Tw}.

Parry and Evans \cite{pe1} used scaling hypothesis to postulate that 
for $M\to\infty$
\begin{equation}
\Tw-\TwM \sim M^{-1/\beta_s}.
\end{equation}
Because for a 2D Ising model $\beta_s=1$ the behaviour of the difference between pseudotransition temperature $\TwMG$ and $\Tw$ agrees with this hypothesis.

Similarly, for $\TcMG$ one obtains from \eref{TwM:TcM}
\begin{equation}
\frac{\TcMG-\Tc}{\Tc}=\frac{A_2}{M}+\Or\left(M^{-2}\right),
\end{equation} 
where the amplitude $A_2=\left[2\ln\left(1+\sqrt{2}\right)\right]^{-1}$ is universal. Since for a 2D Ising model one has $\nu=1$, thus $\left(\TcMG-\Tc\right)/\Tc\sim M^{-1/\nu}$ as expected on the basis of scaling arguments \cite{barber}. We note that $\TcMG$ is always larger than $\Tc$.

The wetting temperature is a continuous function of the surface field $h_1$. Parry and Evans \cite{pe1} proposed the scaling function $X_{\rm AS}$ which describes the dependence of $\TwM$ on the width of the strip $M$ and the surface field $h_1$ in the limit $h_1\to 0$ and $M\to \infty$ with $h_1 M^{\Delta_1/\nu}$ fixed 
\begin{equation}\label{scale1}
\frac{\Tc-\TwM}{\Tc}=M^{-1/\nu}X_{\rm AS}\left( h_1 M^{\Delta_1/\nu}\right).
\end{equation}
It turns out that the pseudocritical temperature $\TwM^\gamma$ defined by \eref{TwM:TcM} satisfies a similar scaling relation. We have found the exact expression for the corresponding scaling function $X_{\rm AS}^\gamma$. 

For a 2D Ising model $\Delta_1=\frac{1}{2}$ and the scaled variable takes the form $x=h_1 M^{1/2}$.
In order to find the scaling function $X_{\rm AS}^\gamma\left(x\right)$
\begin{equation}
\label{scale2}\frac{\Tc-\TwM^\gamma}{\Tc}=M^{-1}X_{\rm AS}^\gamma\left(x\right)+\Or\left(M^{-2}\right),
\end{equation}
we introduced in \eref{TwM:TcM} the surface field $h_1=x M^{-1/2}$ and obtained in the scaling limit 
\begin{equation}
X_{\rm AS}^\gamma\left(x\right)=\left[2
      \ln\left(1+\sqrt{2}\right)\right]^{-1}+\frac{1}{4}\left(1+\sqrt{2}\right)\ln\left(1+\sqrt{2}\right)\left(\frac{x}{J}\right)^2.
\end{equation}

It is interesting to note that this result is true only for fixed $x$; the scaling function 
\begin{equation}
X_{\rm AS}^\gamma(x)=\lim_{M\to\infty} M\frac{\Tc-\TwM^\gamma\left(M,x M^{-1/2}\right)}{\Tc}
\end{equation}
 is not a uniform limit.
 
The scaling law \eref{scale2} has finite size corrections of order $M^{-2}$ which are present even for $h_1=0$.


\section{Solvation forces}\label{sec3}
 
\subsection{Definition}
The free energy of the strip per column can be calculated from \eref{free:e1} and \eref{free:e2}. For both S and AS cases it naturally decomposes into the sum of three terms
\begin{equation}
\bar{f}^\alpha\left(T,h_1,M\right)=M f_\mathrm{b}\left(T
  \right)+f^\alpha_\mathrm{s}\left(T,h_1\right)+f^\alpha_\mathrm{int}\left( T, h_1, M
    \right),
\end{equation}
where $\alpha\in\left\{\mathrm{S},\mathrm{AS}\right\}$, $f_\mathrm{b}$ is the bulk free energy density \cite{onsager} equal to 
\begin{eqnarray}
\fl f_\mathrm{b}\left(T\right) =-\kb T\left[ \frac{1}{2\pi}\int_0^\pi \arccosh
  \left[ \cosh \left(2K - 2\Kast\right)+1-\cos \omega
  \right] \dd \omega \right.\nonumber \\
   \left.+\frac{1}{2}\ln\left(2\sinh 2K\right)\right],
\end{eqnarray}
$f^\alpha_\mathrm{s}\left(T, h_1\right)$ is the surface free energy per column, and the remaining term
$f^\alpha_\mathrm{int}\left(T,h_1,M\right)$
describes the interaction between the boundaries of the strip per column. By definition, the surface free energy $f^\alpha_\mathrm{s}\left(T,h_1\right)$ does not depend on $M$, and $f^\alpha_\mathrm{int}\left(T,h_1,M\right)$ tends to 0 as $M\to\infty$.

In general, the solvation force is defined as minus derivative of $f^\alpha_\mathrm{int}$ with respect to the distance between the boundary walls. In the present case, because $M$ is integer, we use the definition
\begin{equation}\label{fsolvdef} 
\fsolv^\alpha\left(T, h_1,M\right)=-\left[ f^\alpha_\mathrm{int}\left(T, h_1,
    M+1\right)-f^\alpha_\mathrm{int} \left(T, h_1,M\right)\right]/\kb T ,
\end{equation}
where the factor $1/\kb T$ is additionally introduced to make the solvation force dimensionless. This definition is equivalent to
\begin{equation}\label{fsolvdefuse}
\fl\fsolv^\alpha\left(T,h_1,M\right)=\left[\bar{f}^\alpha\left(T,h_1,M\right)-\bar{f}^\alpha\left(T,h_1,M+1\right)+f_b\left(T\right)\right]/\kb T .
\end{equation}

It is also useful to introduce the difference between the solvation forces corresponding to different boundary fields configurations
\begin{equation}\label{deltafsolv}
\Delta \fsolv \left( T, h_1, M\right)=\fsolv^{\mathrm{AS}}\left( T, h_1, M\right)-\fsolv^{\mathrm{S}}\left( T, h_1, M\right).
\end{equation}
Using \eref{free:e1}, \eref{free:e2} and \eref{fsolvdefuse} it is straightforward to show that
\begin{equation}\label{deltafsolvgamma}
\Delta \fsolv \left( T, h_1, M\right)=\gamma_1\left(T, h_1,M\right)-\gamma_1\left(T, h_1,M+1\right).
\end{equation}
This difference is easier to study analytically than the expression
for $\fsolv^\alpha\left(T,h_1,M\right)$, see \eref{fsolvdefuse}.


\subsection{Basic properties}

We start our analysis by evaluating numerically the solvation forces for different temperatures $T$, strip widths $M$ and surface fields $h_1$.

\begin{figure}
\begin{center}
\begin{tabular}{ll}
(a) & (b) \\
\includegraphics[width=0.45\textwidth]{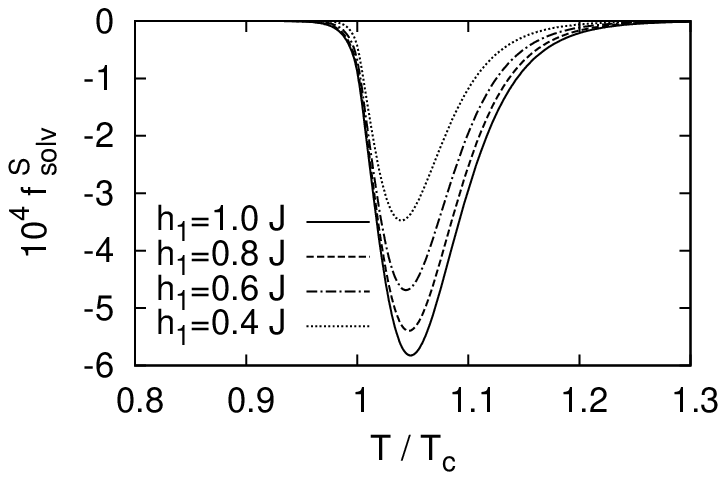} & 
\includegraphics[width=0.45\textwidth]{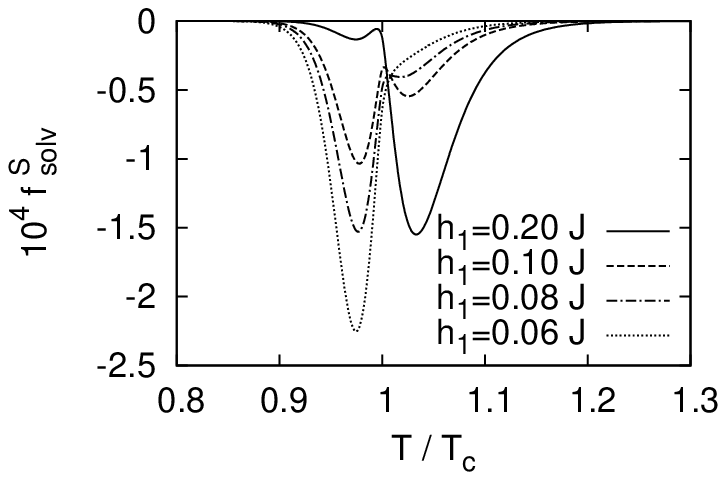} \\
\end{tabular}
\end{center}
\caption{\label{figfsolvS} Plots of the solvation force in the symmetric case ($h_1=h_2$) as a function of temperature for $M=25$ and different values of the boundary field $h_1$.}
\end{figure}

In the symmetric case (S) the solvation force is always negative (attractive). For $h_1$ close to $J$ this force has a minimum at 
$T_\mathrm{min}^{\mathrm{S}>} >\Tc$ and tends to 0 both in the small and large temperature limits. Upon decreasing the boundary field $h_1$, the absolute value of solvation force decreases, and  for $h_1$ small enough a second minimum appears at $T_\mathrm{min}^{\mathrm{S}<} <\Tc$. Upon further decreasing $h_1$, the minimum located at $T_\mathrm{min}^{\mathrm{S}>}$ disappears. The range of $h_1$ for which $\fsolv^\mathrm{S}$ has two minima depends on $M$, and for $M\to\infty$ this range shrinks to $0$. Plots of the solvation force in the symmetric case as a function of temperature for different boundary fields are presented in \fref{figfsolvS}. The behaviour of this force will be studied in detail using scaling functions later on.

\begin{figure}
\begin{center}
\begin{tabular}{ll}
(a) & (b) \\
\includegraphics[width=0.45\textwidth]{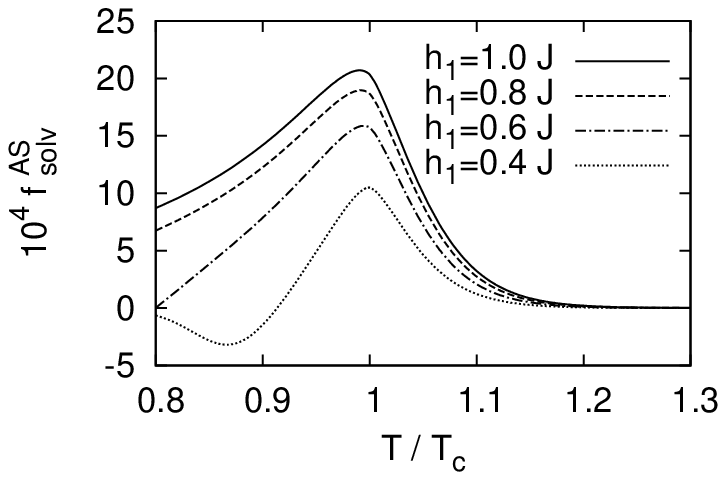} & 
\includegraphics[width=0.45\textwidth]{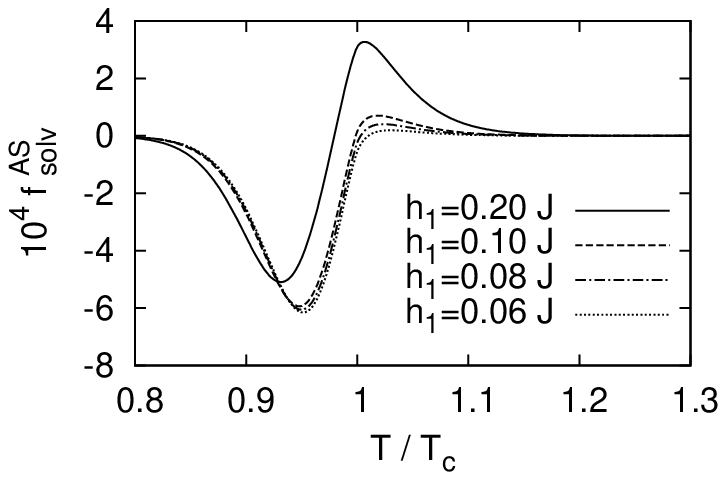} \\
\end{tabular}
\end{center}
\caption{\label{figfsolvAS} Plots of the solvation force in the antisymmetric case ($h_2=-h_1$) as a function of temperature for $M=25$ and different values of the boundary field $h_1$.}
\end{figure}

In the antisymmetric case (AS) the solvation force is plotted in
\fref{figfsolvAS}. For $h_1=J$ this force is positive (repulsive) for
all temperatures and has maximum at $T_\mathrm{max}^{AS}$ located
slightly below $\Tc$. The solvation force
$\fsolv^\mathrm{AS}\left(T,h_1=J,M\right)$ tends to 0 in the high and low
temperature limits. However, for $h_1<J$ the solvation force changes sign. It is negative for
small temperatures, has a minimum at $T_\mathrm{min}^{AS}<\Tw$, and
zero at $\Tast$ slightly above $\Tw$. For temperatures higher than
$\Tast$ the solvation force is positive and has a maximum close to
$\Tc$. For $h_1$ approaching 0, $\Tast$ tends to $\Tc$ and the (negative) value at the minimum
below the wetting temperature decreases. The (positive) maximum value
of the solvation force also decreases and disappears in the limit $h_{1} \to 0$. We
also looked at the location of the maximum of the solvation force
$T_\mathrm{max}^{AS}$. For small $M$, $T_\mathrm{max}^\mathrm{AS}$ is
located above $\Tc$. Upon increasing $M$ the temperature
$T_\mathrm{max}^{AS}$ first crosses the critical temperature and then,
upon further increasing of $M$, approaches $\Tc$ from below. The exact
value of $M$ at which $T_\mathrm{max}^{AS}$ is equal to $\Tc$ depends
on the boundary field $h_1$. We note that the limiting value of the
solvation force at $h_1=0$ is the same for both boundary fields
configurations. Some of the above described properties of solvation force have been
reported for a different system in \cite{T7}.  

The leading $M$-dependence of the solvation force evaluated at $\Tc$ is known exactly \cite{cardy,NI}
\numparts
\begin{eqnarray}
f_\mathrm{solv}^\mathrm{S}\left(\Tc, h_1, M\right) & =& -\frac{\pi}{48 M^2}+\Or\left(1/M^3\right) ,\label{univampa}\\
f_\mathrm{solv}^\mathrm{AS}\left(\Tc, h_1, M\right) & =&\frac{23\pi}{48 M^2}+\Or\left(1/M^3\right),\label{univampb} \\ 
f_\mathrm{solv}^\mathrm{O}\left(\Tc,0,M\right) &=& -\frac{\pi}{48 M^2}+\Or\left(1/M^3\right). \label{univampc}
\end{eqnarray}
\endnumparts
The above values of the universal amplitudes are also recovered numerically in our analysis.

We checked numerically that for $T\neq \Tc$ \cite{ES} 
\begin{equation}\label{fsolvasympt}
\fsolv^\mathrm{S}\left(T,h_1,M\right) \sim \exp\left[-M/\xi_\mathrm{b}\left(T\right)\right],
\end{equation}
where \cite{palmer}
\begin{equation}\label{xib}
\xi_\mathrm{b}=\cases{\left(4K-4\Kast\right)^{-1}& for $T<\Tc$,\\ \left(2\Kast-2K\right)^{-1} & for $T>\Tc$ \\}
\end{equation} 
is the bulk correlation length. Using \eref{deltafsolvgamma} and the dependence of $\gamma_1$ on $M$ for fixed $T$ \cite{ms} we checked that \eref{fsolvasympt} implies the following leading order decay of the solvation force in the antisymmetric case
\begin{equation}
\fsolv^\mathrm{AS}\left(T,h_1,M\right)\sim\cases{\exp\left[-M \ln W\left(T,h_1\right)\right]& for $T<\Tw$,\\ 
                                                 \exp\left[-M/\xi_\mathrm{b}\left(T\right)\right]& for $T=\Tw$,\\ 
                                                 1/M^3 & for $\Tw<T<\Tc$,\\ 
                                                 1/M^2 & for $T=\Tc$,\\ 
                                                 \exp\left[-M/\xi_\mathrm{b}\left(T\right)\right]& for $T>\Tc$.\\}
\end{equation}
The solvation force is a continuous function of temperature and the above formula is correct only in the $M\to\infty$ limit. Below we discuss the behaviour of the solvation force around $\Tw$ and $\Tc$ by introducing the appropriate scaling functions.


\subsection{Scaling at $\Tw$}

To study properties of the solvation force close to $\Tw$ in the antisymmetric case we take the scaling limit $M\to \infty$, $T\to\Tw$ with parameter
$X=M\ln W\left(T,h_1\right) \sim \left(\Tw-T\right)M$ fixed. The function $W\left(T,h_1\right)$
has been introduced in the scaling variable to simplify the scaling function.

To study the solvation force in this limit we use \eref{deltafsolvgamma}. For $T<\Tw$ coefficient $\gamma_1$ is given by \eref{eq:omega1} with $k=1$ and $\alpha_1=-1$
\begin{equation}\label{wettu}
\rme^{-u M}=\rme^{\rmi \delta^\prime\left(\rmi u,T\right)}\frac{W\left(T,h_1,M\right)\rme^{-u}\left(\rme^{-u}-W\left(T,h_1,M\right)^{-1}\right)}{\rme^{-u}-W\left(T,h_1,M\right)},
\end{equation}
where the solution $u$ gives $\gamma_1$ using \eref{omega2gamma} with $\omega_1=\rmi u$.

We put $M=X/\ln W$ in \eref{wettu} and calculate the limit $T\to \Tw$ using l'H\^opital's rule. After introducing $H\left(X\right)=\left.\frac{\partial }{\partial W}u\left(W,X\right)\right|_{W=1} $ one gets 
\begin{equation}\label{H}
\rme^{-X H\left(X\right)}=\frac{H\left(X\right)-1}{H\left(X\right)+1}, \qquad H\left(X\right)<0.
\end{equation}
The function $H\left(X\right)$ can be calculated numerically for any $X>0$. The solution of \eref{wettu} in the scaling limit takes the following form
\begin{equation}
u=H\left(X\right) \ln W+\Or\left(\ln^2 W\right),
\end{equation}
and one obtains\footnote{Note that the expression \eref{gamma1X} for the function $\gamma_1\left(X,h_1,M\right)$ is not equivalent to equation (4.7) in \cite{maciolek}, because of an error in calculation.}
\begin{equation}\label{gamma1X}
\gamma_1\left(X,h_1,M\right)=\nu\left(h_1\right)-\frac{1}{M^2}\frac{X^2 H^2\left( X\right)}{2 \sinh\nu\left(h_1\right)}+\Or\left(\frac{1}{M^3}\right),
\end{equation}
where $\nu\left(h_1\right)=\left. (2K-2\Kast) \right|_{T=\Tw\left(h_1\right)}$.

Comparison of this result with \eref{fsolvasympt} leads to the conclusion that the first term on the right-hand side of \eref{deltafsolv} dominates in the scaling regime (except of $T=\Tw$) and the second term may be neglected.

With the help of \eref{gamma1X} one gets 
\begin{equation}\label{fsolvscaleTw}
f_\mathrm{solv}^\mathrm{AS}\left(X,h_1,M\right)=-\frac{1}{M^3} G\left(X,h_1\right)+\Or\left(\frac{1}{M^4}\right),
\end{equation}
where
\begin{equation}\label{G}
G\left( X,h_1\right)=\frac{X^3 H^2\left(X\right)}{\sinh \nu\left(h_1\right)}\ \frac{H^2\left(X\right)-1}{2+X\left(H^2\left(X\right)-1\right)}.
\end{equation}

\begin{figure}
\begin{center}
\includegraphics[width=0.45\textwidth]{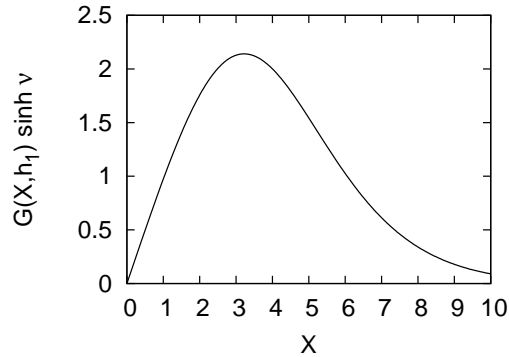}
\end{center}
\caption{\label{figG}The scaling function $G\left(X,h_1\right)$ for the solvation force in the antisymmetric case multiplied by $\sinh \nu\left( h_1\right)$ (see \eref{fsolvscaleTw}).}
\end{figure}

Using \eref{H} and \eref{G} it is straightforward to analyse the properties of the scaling function $G\left(X,h_1\right)$. For small $X$
\begin{equation}\label{GsmallX}
G\left(X,h_1\right)=\frac{X}{\sinh \nu}+\Or\left(X^2\right),
\end{equation} 
it has a maximum at $X_0\approx 3.22149$ and approaches zero exponentially for large $X$, see \fref{figG}.

The above properties can be used to explain the behaviour of the solvation force around the wetting temperature for large $M$. In particular, one has 
\begin{equation}
\frac{\Tw-T_\mathrm{min}^\mathrm{AS}}{\Tc}=\frac{A_3\left(h_1\right)}{M}+\Or\left(\frac{1}{M^2}\right),
\end{equation}
\begin{equation}
f_\mathrm{solv}^\mathrm{AS}\left(T_\mathrm{min}^\mathrm{AS},h_1,M\right)=-\frac{A_4\left(h_1\right)}{M^3}+\Or\left(\frac{1}{M^4}\right),
\end{equation}
where the functions $A_3\left(h_1\right)$ and
$A_4\left(h_1\right)$ are positive and may be obtained from the scaling function $G$
and the definition of scaling variable $X$.

The dependence of $T^\ast$ on $M$ can be explained using
\eref{deltafsolv}. Exactly at $T=\Tw$ the left-hand side of this equation
is zero (exactly at $\Tw$ the coefficient $\gamma_1\left(T,h_1,M\right)$ does not depend on $M$) so the solvation force is the same for both AS and S boundary fields. With the help of equations  \eref{deltafsolvgamma}, \eref{fsolvasympt} and
\eref{GsmallX} one gets 
\begin{equation}
\frac{T^\ast-\Tw}{\Tc}=A_5\left(h_1\right)M^2\exp\left[-M/\xi_\mathrm{b}\left(\Tw\left(h_1\right)\right)\right],
\end{equation}
where $A_5\left(h_1\right)$ is a positive function. 

\subsection{Scaling at $\Tc$}

For temperatures close to the bulk critical temperature the solvation force takes in the limit $T\to \Tc$, $M\to\infty$ with fixed $h_1$ and $\bar{x}=\mbox{sign}\left(T-\Tc\right) M/\xi_b\left(T\right)$, the following scaling form 
\begin{equation}\label{xfsolv}
f_\mathrm{solv}^\mathrm{\alpha}\left(T,h_1,M\right)=\frac{1}{M^2} \mathcal{X}^\alpha_{h_1}\left(\bar{x}\right)+\Or\left(M^{-3}\right).
\end{equation}
Note that the factor $\sign\left(T-\Tc\right)$ introduced in the definition of the scaling variable $\bar{x}$ makes it negative for $T<\Tc$ and positive for $T>\Tc$. The bulk correlation length \eref{xib} close to $\Tc$ takes the form 
\begin{equation}
\xi_b\left(T\right)\approx \cases{\xi_0^+ \left|t\right|^{-1} & for $T>\Tc$ ($x>0$),\\
                                  \xi_0^- \left|t\right|^{-1} & for $T<\Tc$ ($x<0$),\\} 
\end{equation}
where $t=\left(T-\Tc\right)/\Tc$ and the amplitudes $\xi_0^+=2\xi_0^-=1/\left[2\ln\left(1+\sqrt{2}\right)\right],$ such that $\bar{x}\sim \left(T-\Tc\right) M$ for $T$ close to $\Tc$. Later on we will use 
\begin{equation}
x= t M/\xi_0^+ \approx \cases{\bar{x} & for $\bar{x}>0$, \\ \bar{x}/2 & for $\bar{x}<0$, \\}
\end{equation}
instead of the scaling variable $\bar{x}$.

The scaling function $\mathcal{X}$ has already been proposed by Evans and Stecki \cite{ES} and has been calculated analytically in both S and AS cases for particular value of the scaling field $h_1=J$. Here we consider arbitrary values of $h_1$. Our numerical calculations show that, up to numerical errors,
\begin{equation}\label{Xuniversal}
\mathcal{X}^\alpha_{h_1}\left(x\right)=\mathcal{X}^\alpha_J\left(x\right) \quad\mbox{for}\quad h_1\neq 0.
\end{equation}
We have found numerically that the corrections to scaling depend on $h_1$ and are getting smaller for $h_1$ close to $J$. Unfortunately, we are unable to prove analytically this property of scaling function
$\mathcal{X}_{h_1}^{\alpha}\left(x\right)$.

However, the difference between the scaling functions for the two cases
\begin{equation}
\Delta \mathcal{X}_{h_1}\left(x\right)=\mathcal{X}^\mathrm{AS}_{h_1}\left(x\right)-\mathcal{X}^\mathrm{S}_{h_1}\left(x\right)
\end{equation}
can be calculated exactly.  For $T$ close to $\Tc$ this function is obtained from  \eref{deltafsolvgamma}, \eref{omega2gamma}, \eref{eq:omega} and \eref{l}. To derive $\omega_1$ in the scaling limit we replace $M$ with $x \xi_0^+/t$ in \eref{eq:omega} and use the following property of the function $\phi\left(\omega,T,h_1\right)$ 
\begin{equation}
\lim_{T\to\Tc} \phi\left(\omega_1\left(T,h_1,M\right),T,h_1\right)=0
\end{equation}
for $h_1\neq 0$. Thus the only term that depends on the surface field in \eref{eq:omega} disappears in this scaling limit and the calculation of $\Delta \mathcal{X}$ goes along the same lines as in \cite{ES} (from now on we drop the index $h_1$ in $\Delta \mathcal{X}$). This is in full agreement with \eref{Xuniversal}. One obtains
\begin{equation}
\Delta \mathcal{X}=\frac{w^2 \sin w}{w-\sin w \cos w},
\end{equation}
with $w$ being a solution of 
\begin{equation}
w \cot w=x,
\end{equation}
where $0\leq w < \pi$ for $x\leq 1$, and $w=\rmi u,$ $u>0$ for $x>1$.

\begin{figure}
\begin{center}
\begin{tabular}{ll}
(a) & (b) \\
\includegraphics[width=0.45\textwidth]{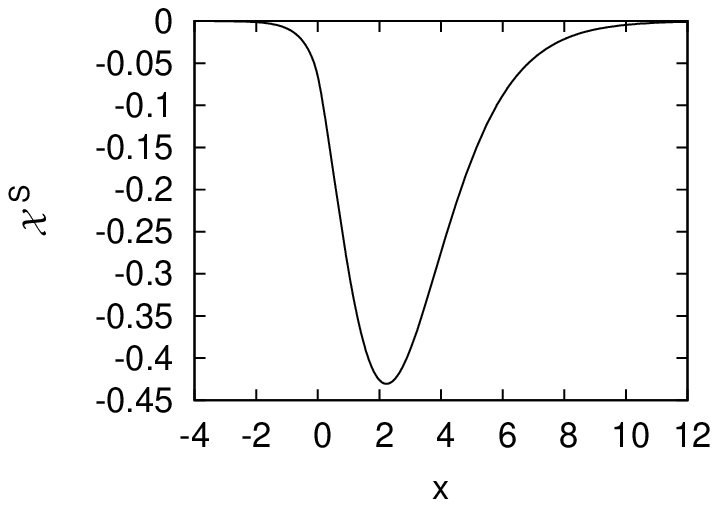} & 
\includegraphics[width=0.45\textwidth]{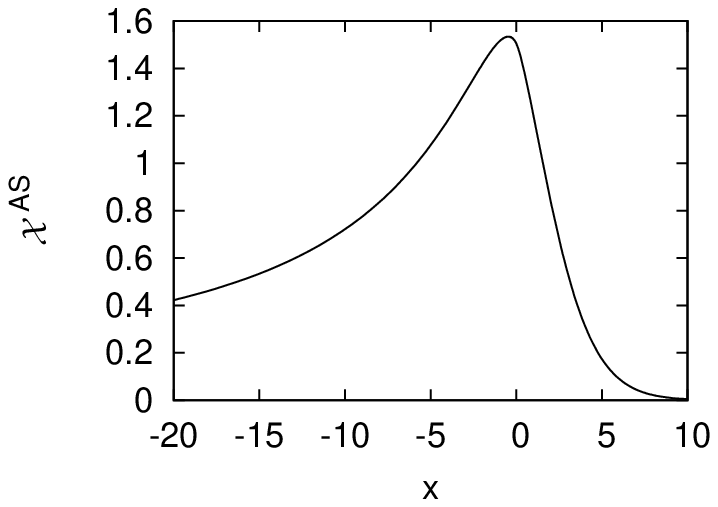} \\
\end{tabular}
\end{center}
\caption{\label{figX} The scaling function $\mathcal{X}$ describing the solvation force for any $h_1\neq 0$ and $M\to\infty$ with $x=t M/\xi_0^+$ fixed for: (a) symmetric ($h_1=h_2$), and (b) antisymmetric ($h_1=-h_2$) boundary fields. Each plot does not depend on the chosen value of $h_1$ and is the same (up to numerical errors smaller than the resolution of the plot) as the analytically calculated scaling functions for $h_1=J$ \cite{ES}.}
\end{figure}

Function $\mathcal{X}_{h_1}^\mathrm{S}\left(x\right)$ for different $h_1$-values
is plotted in \fref{figX}a. Note that such obtained curves are indistinguishable from each other which numerically proves \eref{Xuniversal} for the symmetric case. It has a minimum for $x>0$, so from \eref{xfsolv} it follows that for large $M$ 
\begin{eqnarray}
T_\mathrm{min}^{\mathrm{S}>}\left(h_1,M\right)&=&\Tc\left[1+\frac{A_6}{M}+\Or\left(M^{-2}\right)\right],\\
\fsolv^\mathrm{S}\left(T_\mathrm{min}^{\mathrm{S}>},h_1,M\right) & =& -\frac{A_7}{M^2}+\Or\left(M^{-3}\right),
\end{eqnarray}
with $A_6$ and $A_7$ determined by the position of minimum of the scaling function
\begin{equation}
A_6\approx 1.26424, \qquad A_7\approx0.43052.
\end{equation}
Because $\mathcal{X}_{h_1}^\mathrm{S}\left(x\right)$ has only one minimum, the second minimum of the solvation force, located below $\Tc$, disappears in this limit.

Function $\mathcal{X}_{h_1}^\mathrm{AS}\left(x\right)$ for for different $h_1$-values 
 is plotted in \fref{figX}b. Again one notes that such obtained curves are indistinguishable from each other which numerically proves \eref{Xuniversal} for the antisymmetric case. It has a maximum for $x<0$ and from \eref{xfsolv} it follows that for large $M$
\begin{eqnarray}
T_\mathrm{max}^\mathrm{AS}\left(h_1,M\right)&=&\Tc\left[1-\frac{A_8}{M}+\Or\left(M^{-2}\right)\right],\\
\fsolv^\mathrm{AS}\left(T_\mathrm{max}^\mathrm{AS},h_1,M\right) & =& \frac{A_9}{M^2}+\Or\left(M^{-3}\right),
\end{eqnarray}
with $A_8$ and $A_9$ determined by the position of maximum of the scaling function
\begin{equation}\label{A89}
A_8\approx0.2651, \qquad A_9\approx1.5341.
\end{equation}
The temperature $T_\mathrm{max}^\mathrm{AS}$ is smaller than $\Tc$ in
this limit. These results have already been reported in \cite{ES} for
$h_1=J$. According to our numerical analysis the values of constants
$A_6, A_7, A_8$ and $A_9$ are the same for any nonzero surface field
$h_1$. \footnote{Note a minor disagreement between values of our numerical
  amplitudes $A_8$ and $A_9$ \eref{A89} and those evaluated in
  \cite{ES} due to minor numerical inaccuracies in \cite{ES}.} 

\subsection{Scaling for $\Tw\to\Tc$}

To explain the properties of the solvation force for small values of the boundary field $h_1$ we consider the scaling limit $M\to \infty$, $T\to\Tc$ and $h_1\to 0$ (i.e. $\Tw\to\Tc$) with two scaling variables
\begin{equation}\label{Yscalevar}
x=\frac{t M}{\xi_0^+}, \qquad y=\frac{A_0}{\kb \Tc} \frac{h_1}{\left|t\right|^{1/2}} 
\end{equation}
fixed. In this limit the solvation force can be described by scaling function $\mathcal{Y}^\alpha\left(x,y\right)$
\begin{equation}\label{Yscale}
\fsolv^{\alpha}\left(T, h_1, M\right)=\frac{1}{M^2} \mathcal{Y}^{\alpha}\left(x,y\right)+\Or\left(M^{-3}\right).
\end{equation}
The constant
$A_0=\left[\left(1+\sqrt{2}\right)/\ln\left(1+\sqrt{2}\right)\right]^{1/2}$
in \eref{Yscalevar} was chosen such that for negative values of $x$, the value
$y=1$ corresponds to $T=\Tw$. For $y<1$ equation \eref{Yscale} gives
the solvation force for $T$ below $\Tw$, and for $y>1$ -- for $T$ above
$\Tw$. This scaling function has already been analysed for subcritical temperatures in \cite{NN}.\footnote{There is a mistake in the scale of variable $x$ in figures 3, 4 and 5 in \cite{NN}. To get the correct values of $x$ one should replace $x$ by $\left(\xi_0^-\right)^{-2} x$ in these figures in \cite{NN}.} 

The scaling function $\mathcal{Y}^{\alpha}\left(x,y\right)$ can only be calculated numerically; details of evaluation are presented in \ref{appendix}. 

Before presenting the numerically evaluated properties of the scaling functions $\mathcal{Y}^\alpha\left(x,y\right)$, $\alpha=\mathrm{S}, \mathrm{AS}$ we note that one can test some of these properties through analytically determined difference 
\begin{equation}\label{DeltaY}
\Delta \mathcal{Y}\left(x,y\right)=\mathcal{Y}^\mathrm{AS}\left(x,y\right)-\mathcal{Y}^{S}\left(x,y\right).
\end{equation}
This can be done with the help of \eref{deltafsolvgamma}. The coefficient $\gamma_1$ is given by equation \eref{eq:omega},
where its solution $\omega_1$ determines  $\gamma_1$ by \eref{omega2gamma}. After applying the scaling limit to the above equation one gets
\begin{equation}
\gamma_1\left(T,h_1,M\right)=\frac{1}{M} \sqrt{x^2+w^2}+\Or\left(M^{-2}\right),
\end{equation}
where $w$ is a solution of
\begin{equation}\label{wY}
w\cot w=x \frac{x^2\left[y^2+\sign\left(x\right)\right]^2+w^2\left[1+2 \sign\left(x\right) y^2\right]}{x^2\left(y^4-1\right)-w^2}.
\end{equation}
Depending on $x$ and $y$ equation \eref{wY} may have many different solutions for $w$. The rules for choosing the correct solution are summarised in \tref{table1}; other solutions give the coefficients $\gamma_k$ for $k>1$.

\begin{table}
\caption{\label{table1} The rules for picking the correct solutions of \eref{wY} corresponding to $\gamma_1$. In all four cases there exists exactly one solution for a given range of $w$.}
\begin{indented}
\item[]\begin{tabular}{@{}lll}
\br
Range of $x$ and $y$ & Domain of $w$ & Condition on $w$\\
\mr
$y>1, x\leq\left(y^2-1\right)/\left(y^2+1\right)$ & real & $0\leq w<\min\left\{\pi,\left|x\right|\left(y^4-1\right)^{1/2}\right\}$ \\
$y>1, x>\left(y^2-1\right)/\left(y^2+1\right)$ & imaginary & $0<w/\rmi<x\left(y^2+1\right)\left(1+2y^2\right)^{-1/2}$\\
$y\leq 1, x>0$ & imaginary & $x \left(1-y^4\right)^{1/2}<w/\rmi$\\
& & $w/\rmi<x\left(y^2+1\right)\left(1+2y^2\right)^{-1/2}$\\
$y\leq 1, x\leq 0$ & imaginary & $\left|x\right| \left(1-y^2\right)\leq w/\rmi\leq \left|x\right| \left(1-y^4\right)^{1/2}$\\
\br
\end{tabular}
\end{indented}
\end{table}

The function $\Delta \mathcal{Y}$ is given by the formula
\begin{equation}
\Delta \mathcal{Y}\left(x,y\right)=-\lim_{M\to \infty}M^2 \left(\frac{\partial \gamma_1}{\partial M}\right)_{T,h_1},
\end{equation}
which leads to a rather lengthy expression and we refrain from presenting it here. 

\begin{figure}
\begin{center}
\begin{tabular}{ll}
(a) & (b) \\
\includegraphics[width=0.45\textwidth]{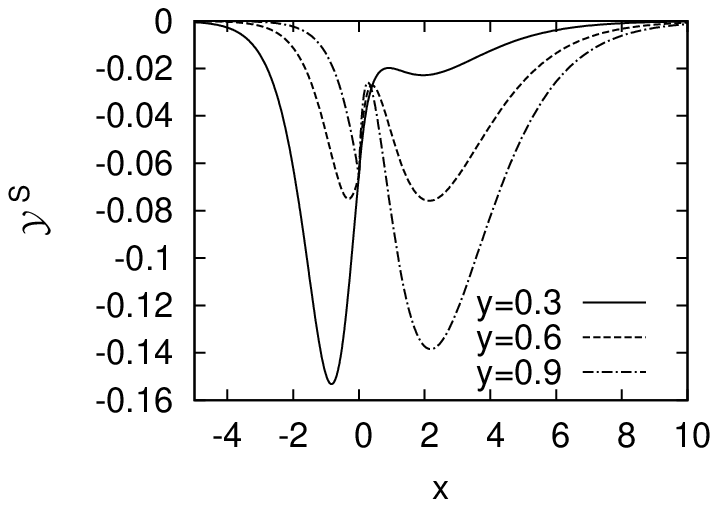} & 
\includegraphics[width=0.45\textwidth]{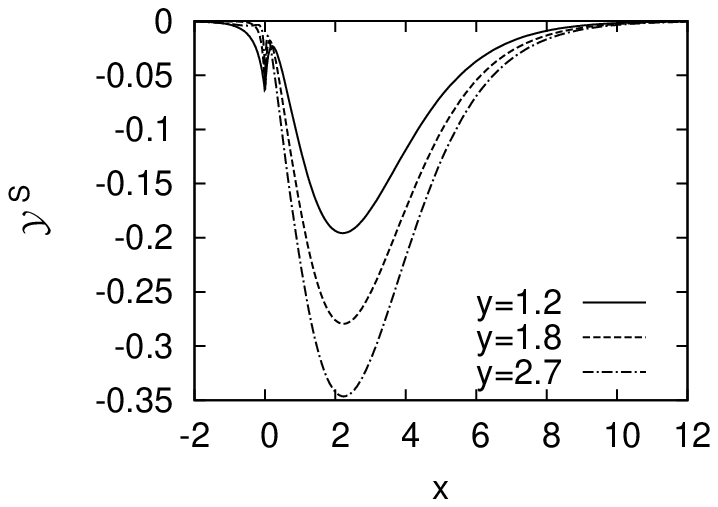} \\
(c) & (d) \\
\includegraphics[width=0.45\textwidth]{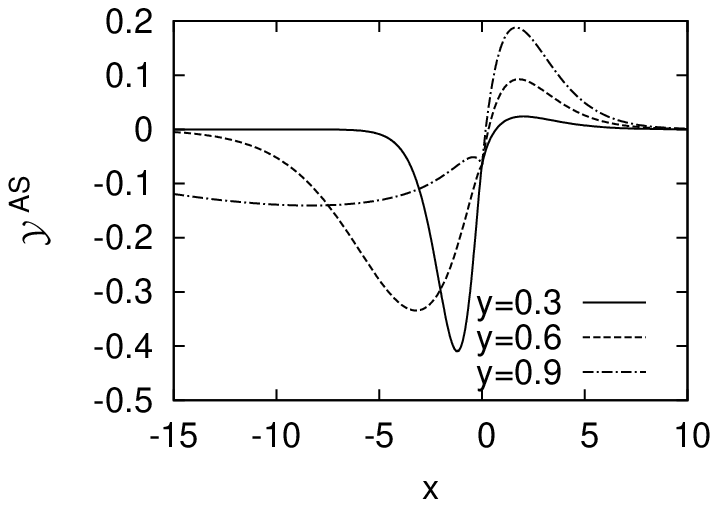} & 
\includegraphics[width=0.45\textwidth]{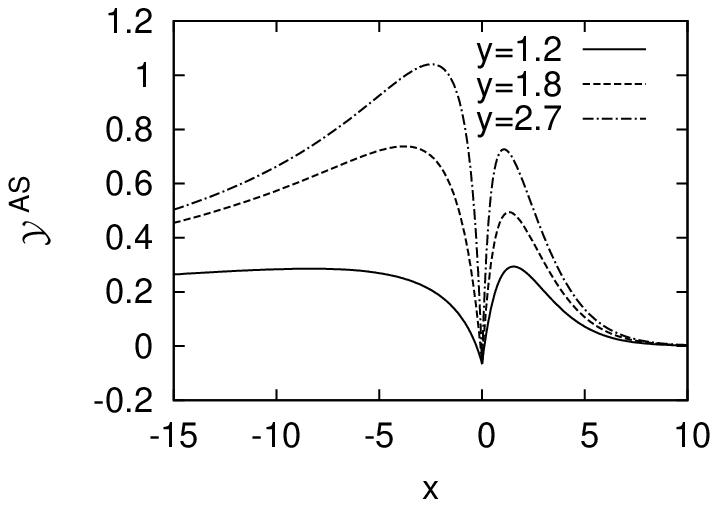} \\
\end{tabular}
\end{center}
\caption{\label{figY} The scaling function $\mathcal{Y}^\alpha\left(x,y\right)$ as function of $x$ for different values of $y$. Boundary fields are symmetric ($\alpha=\mathrm{S}$) in graphs (a) and (b), and antisymmetric ($\alpha=\mathrm{AS}$) in graphs (c) and (d).}
\end{figure}

The scaling functions $\mathcal{Y}^\alpha\left(x,y\right)$ are plotted in \fref{figY}. These plots cannot be used directly to approximate the behaviour of the solvation force as a function of temperature for large fixed $M$, because for fixed $y$ both the  temperature $T$ and the surface field $h_1$ become functions of $x$. Additionally, the limit $x\to 0$ corresponds to $h_1 \to 0$, which explains why --- for any $y$ --- one has
\begin{equation}
\mathcal{Y}^{\mathrm{AS}}\left(0,y\right)=\mathcal{Y}^{\mathrm{S}}\left(0,y\right)=-\frac{\pi}{48},
\end{equation}
i.e. in this limit the scaling function equals the universal amplitude describing the decay of the solvation force at $T=\Tc$ for free boundary conditions.

\begin{figure}
\begin{center}
\begin{tabular}{ll}
(a) & (b) \\
\includegraphics[width=0.45\textwidth]{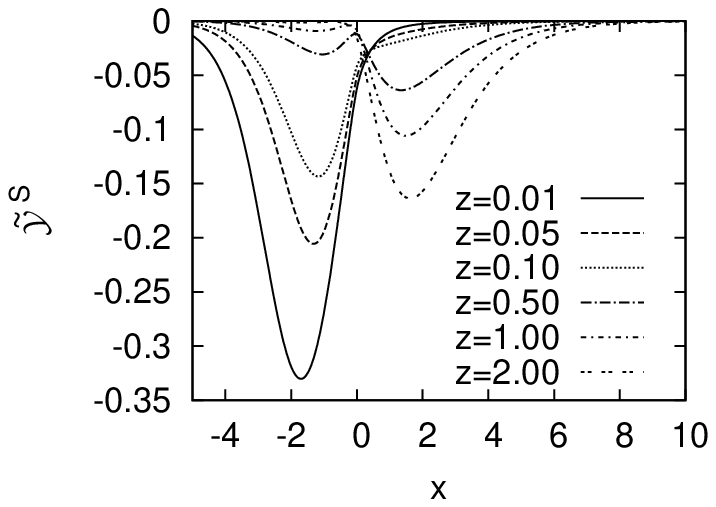} & 
\includegraphics[width=0.45\textwidth]{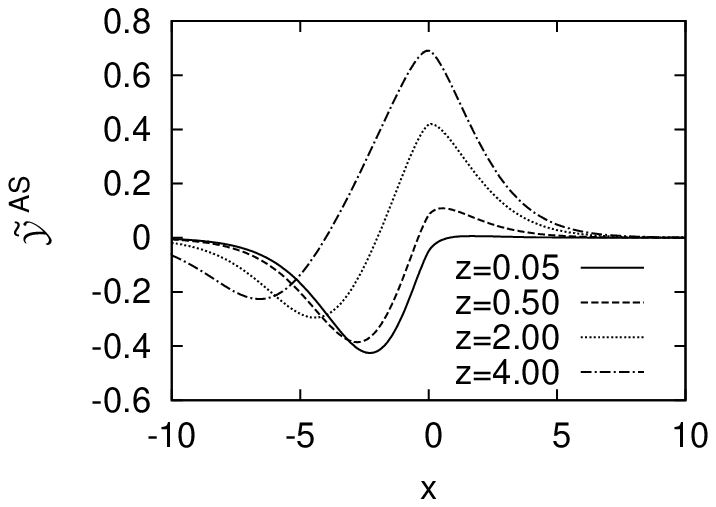} \\
\end{tabular}
\end{center}
\caption{\label{figYz} The scaling function $\tilde\mathcal{Y}^\alpha\left(x,y\right)$ for fixed $z$ as a function of $x$. Boundary fields are symmetric ($\alpha=\mathrm{S}$) in graph (a) and antisymmetric ($\alpha=\mathrm{AS}$) in graph (b).}
\end{figure}

To explain the observed properties of solvation force we changed variables in the scaling function and defined new function
\begin{equation}
\tilde{\mathcal{Y}}^\alpha\left(x,z\right)=\mathcal{Y}^\alpha\left(x,\sqrt{z/\left|x\right|}\right),
\end{equation}
where the new variable $z=\left|x\right| y^2\sim M h_1^2$, so that fixing $x$ and $z$ is equivalent to fixing $x$ and $y$ in the scaling limit. For the new scaling variables one obtains
\begin{equation}\label{Yscalez}
\fsolv^\alpha\left(T,h_1,M\right)=\frac{1}{M^2}\tilde\mathcal{Y}^\alpha\left(x,z\right)+\Or\left(1/M^3\right),
\end{equation}
which can be used to approximate the solvation force as a function of temperature for fixed $M$ and $h_1$. Plots of $\tilde \mathcal{Y}$ are presented in \fref{figYz}.

We checked that up to our numerical precision
\begin{equation}
\lim_{z\to\infty}\tilde\mathcal{Y}^\alpha\left(x,z\right)=\mathcal{X}^\alpha\left(x\right),\qquad \lim_{z\to 0}\tilde\mathcal{Y}^\alpha\left(x,z\right)=\mathcal{X}^0\left(x\right),
\end{equation}
which is not surprising since for $M\to\infty$ and fixed $h_1\neq 0$ one has $z\sim Mh_1^2\to\infty$, while $h_1=0$ implies $z=0$.

In both S and AS cases the scaling functions $\tilde\mathcal{Y}^\alpha\left(x,y\right)$ reflect the behaviour of solvation force for small surface fields. For symmetric surface fields, the negative function $\tilde\mathcal{Y}^\mathrm{S}\left(x,z\right)$ has for fixed $z<z_1\approx0.1474$ only one minimum at negative values of $x$. At $z=z_1$ the second minimum located at positive $x$ appears. Upon further increasing of $z$, the minimum at $x<0$ is increasing and the absolute value of the second minimum is increasing. We could not observe the disappearance of the minimum at negative $x$, because for large $z$ the depth of this minimum is of order of our numerical errors.
 
For antisymmetric surface fields the scaling function $\tilde\mathcal{Y}^\mathrm{AS}\left(x,z\right)$ has exactly one minimum and one maximum for all finite $z$. The minimum is always located at $x<0$ and moves towards $-\infty$ when $z$ is increased. The maximum moves from $x=\infty$ for $z=0$ to a finite negative value of $x$ for $z=\infty$. The value of scaling function at maximum is always positive and becomes very small for $z$ close to $0$. For $z=z_2\approx 0.212$ the scaling function vanishes at $x=0$, which means that (up to higher order corrections) the solvation force disappears at $T=\Tc$. On the other hand, at $z=z_3\approx3.35$, the maximum of scaling function is located exactly at $x=0$. 

The above observations are summed up below:
\begin{itemize}
\item in the S case the minimum of the solvation force located above $\Tc$ exists for 
\begin{equation}
h_1/J>\frac{A_{10}}{\sqrt{M}}+\Or\left(M^{-3/2}\right),\qquad A_{10}\approx 0.40,
\end{equation}
\item in the AS case the solvation force is zero at $T=\Tc$ for
\begin{equation}
h_1/J=\frac{A_{11}}{\sqrt{M}}+\Or\left(M^{-3/2}\right),\qquad A_{11}\approx 0.48,
\end{equation}
\item in the AS case the maximum of the solvation force is located exactly at $T=\Tc$ for
\begin{equation}
h_1/J=\frac{A_{12}}{\sqrt{M}}+\Or\left(M^{-3/2}\right),\qquad A_{12}\approx 1.89.
\end{equation}
\end{itemize}

\begin{figure}
\begin{center}
\begin{tabular}{ll}
(a) & (b) \\
\includegraphics[width=0.45\textwidth]{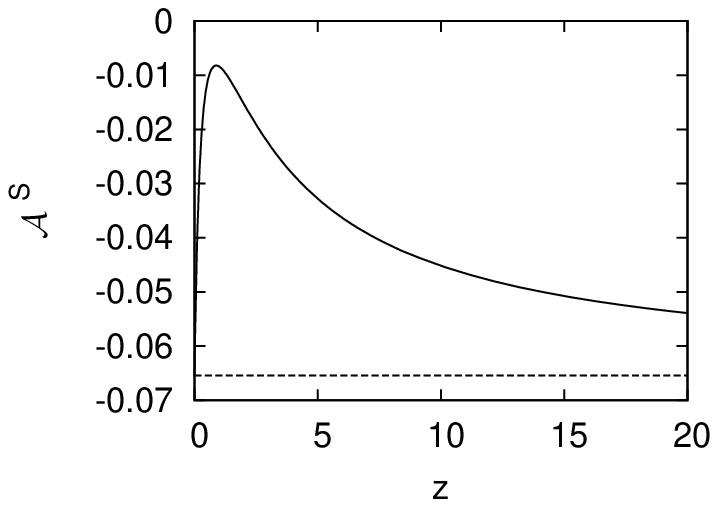} & 
\includegraphics[width=0.45\textwidth]{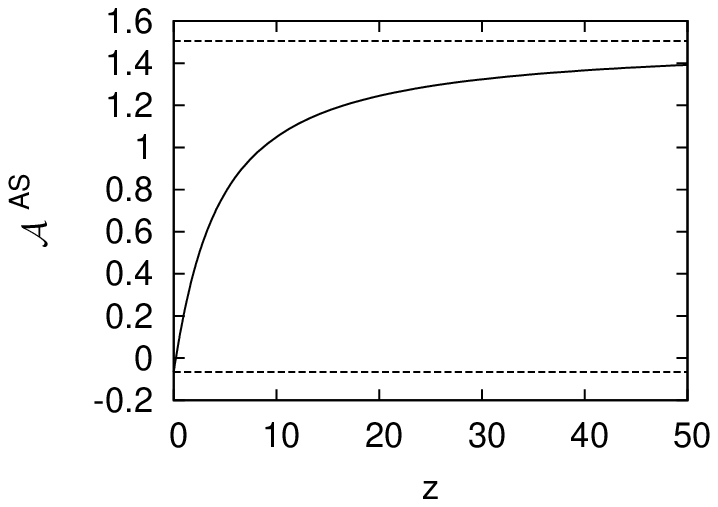} \\
\end{tabular}
\end{center}
\caption{\label{amplitudes}The amplitudes $\mathcal{A}^\alpha\left(z\right)$ describing the decay of the solvation force at $T=\Tc$ for $M\to\infty$ (see \eref{ampA}); (a) symmetric surface fields, (b) antisymmetric surface fields. Dashed lines show the exactly known values of amplitudes for $z=0$ and $z\to\infty$.}
\end{figure}

Finally, we study the behaviour of the solvation force at $T=\Tc$
\begin{equation}\label{ampA}
\fsolv^\alpha\left(\Tc,h_1,M\right)=\frac{\mathcal{A}^\alpha\left(z\right)}{M^2}+\Or\left(M^{-3}\right).
\end{equation}
The amplitude $\mathcal{A}^\alpha\left(z\right)$ can be calculated using the scaling function
\begin{equation}
\mathcal{A}^\alpha\left(z\right)=\tilde\mathcal{Y}^\alpha\left(x=0,z\right).
\end{equation}
From \eref{univampa}, \eref{univampb} and \eref{univampc} the values of the amplitudes $\mathcal{A}^\alpha\left(z\right)$ for $z=0$ and $z\to\infty$ follow
\begin{equation}
\fl\mathcal{A}^\mathrm{S}\left(z=0\right)=\mathcal{A}^\mathrm{AS}\left(z=0\right)=\mathcal{A}^\mathrm{S}\left(z=\infty\right)=-\frac{\pi}{48},\qquad \mathcal{A}^\mathrm{AS}\left(z=\infty\right)=\frac{23 \pi}{48}.
\end{equation}
For other values of $z$ amplitudes $\mathcal{A}^\alpha\left(z\right)$ can be calculated numerically; they are presented in \fref{amplitudes}.

\section{Summary}

In this article we considered the two-dimensional Ising strip of width $M$ with surface fields $h_1$ and $h_2$ acting on the boundaries of the system. We considered only symmetric ($h_1=h_2$) and antisymmetric ($h_1=-h_2$) configurations of the surface fields. 

We introduced two pseudotransition temperatures: $\TwMG$ and
$\TcMG$. Around $\TwMG$, in the antisymmetric case, the interface
separating two magnetic phases moves from position close to one wall
to the centre of the strip. At $\TcMG$ the difference between the two
phases disappears. The existence of these two temperatures follows
from the properties of our solution for the free energy. We proved
that $\TwMG$ and $\TcMG$ have the same scaling properties as real transition
temperatures in higher dimensions. We also checked scaling relations of $\TwMG$ postulated by Parry and Evans \cite{pe1}. 

The major part of our analysis was concentrated on the properties of the solvation force. We calculated this force as a function of temperature $T$, surface field $h_1$ and strip width $M$. For symmetric surface fields this force is always negative (attractive). For strong surface fields the solvation force has a minimum above the bulk critical temperature of the 2D system $\Tc$, while for small surface fields the minimum is located below $\Tc$. There exists a range of surface fields for which this force has two minima. For antisymmetric surface fields (and $h_1 \neq J$) the solvation force 
changes the sign: it is negative for small temperatures and positive (repulsive) for high temperatures. The temperature $\Tast$ at which the solvation force is zero is located very close to the wetting temperature of the semi-infinite system $\Tw$. For large surface fields the solvation force has a maximum below $\Tc$. When the surface field is decreased, this maximum crosses $\Tc$. Upon further decrease of $h_1$ the maximum disappears.

To explain these properties we proposed scaling functions in three different scaling regimes: at $\Tw$, at $\Tc$, and in the case when $\Tw\to\Tc$.

For antisymmetric surface fields close to $\Tw$ we found scaling in the limit $M\to \infty$ and $T\to\Tw$ with $M\left(T-\Tw\right)$ fixed. We succeeded in finding the analytical formula for the scaling function and used it to explain the behaviour of the solvation force. The scaling function is nonuniversal, i.e. it depends on the magnitude of surface field. 

Close to $\Tc$ the scaling limit is $M\to\infty$ and $T\to \Tc$ with
$M\left(T-\Tc\right)$ fixed. For both symmetric and antisymmetric
surface fields the obtained scaling function is, within our numerical
accuracy, independent of $h_1$ for $h_1\neq0$. We also showed analytically
that the difference between scaling functions in both configurations
of surface fields is independent of $h_1$. Using properties of such
obtained scaling functions we explained the location of maxima and minima of the solvation
force around $\Tc$ for strong surface fields. 

The third scaling limit corresponds to $h_1\to 0$,
which implies $\Tw\to\Tc$, $T\to\Tc$ and $M\to \infty$ with
$M\left(T-\Tc\right)$ and $Mh_1^2$ fixed. In this limit we calculated
the scaling function numerically and checked that it explains the location
of minima and maxima of the solvation force for small surface fields. 

\ack

Helpful discussions with A. Macio{\l}ek, D. Danchev, S. Dietrich, O. Vasilyev, and F. Toldin are gratefully acknowledged.

\appendix
\section{Numerical calculation of scaling function $\mathcal{Y}^\alpha\left(x,y\right)$}
\label{appendix}
In this appendix we explain methods used to calculate the scaling functions. From \eref{Yscalevar} and \eref{Yscale} we get the formula
\begin{equation}
\fl\mathcal{Y}^\alpha\left(x,y\right)=\lim_{M\to\infty}\mathcal{Y}^\alpha_M\left(x,y\right), \qquad \mathcal{Y}^\alpha_M\left(x,y\right)=M^2 \fsolv^\alpha\left(T\left(x,M\right),h_1\left(x,y,M\right),M\right),
\end{equation}
where $T\left(x,M\right)=\Tc\left(1+x\xi_0^+/M\right)$ and $h_1\left(x,y,M\right)=y \kb \Tc/A_0\left(x\xi_0^+/M\right)^{1/2}$. Function $\mathcal{Y}^\alpha_M\left(x,y\right)$ can be calculated numerically with arbitrary numerical precision. However, when $M$ is large or high  precision is required, the time spend on calculation becomes very long. Although the limiting value $\mathcal{Y}^\alpha\left(x,y\right)$ cannot be calculated exactly, it may be estimated in several ways. One possibility is to fix a large but finite $M$ and assume
\begin{equation}
\mathcal{Y}^\alpha\left(x,y\right)\approx \mathcal{Y}^\alpha_{M}\left(x,y\right).
\end{equation}
This method was used in \cite{NN} with $M=200$. 

In this paper we applied the least squares method. Because the difference  $\mathcal{Y}^\alpha\left(x,y\right)-\mathcal{Y}_M^\alpha\left(x,y\right)$ depends on $M$ and its absolute values are large for small $M$, this method cannot be used directly. 

To overcome this problem and to estimate the value of $\mathcal{Y}^\alpha\left(x,y\right)$ we calculate the values of $\mathcal{Y}_M^\alpha\left(x,y\right)$ for $M=M_0, M_0+1,M_0+2,\ldots M_0+m$ and fit the results to the formula
\begin{equation}
\mathcal{Y}_M^\alpha\left(x,y\right)=B_0+\frac{B_1}{M}+\frac{B_2}{M^2}+\ldots+\frac{B_n}{M^n},
\end{equation}
which we assume to reflect the form of leading corrections to the scaling. We take $B_0$ as our estimate of $\mathcal{Y}^\alpha\left(x,y\right)$. The accuracy of this algorithm depends on values of parameters $M_0$, $m$ and $n$. The larger values used the more accurate the result is; we used $M_0=190$, $m=10$ and $n=3$.

The accuracy of such obtained results may be estimated by comparing them with the results obtained for different values of $M_0$. In addition, to test our results we used \eref{DeltaY} with $\Delta \mathcal{Y}$ calculated analytically. The obtained relative accuracy is better than $10^{-4}$.


\end{document}